%% file: main.tex
\documentclass[%
aps,
amsmath,amssymb,floatfix,
reprint,%
twocolumn,%
longbibliography,
reprint,%
author-year,%
author-numerical,%
]{revtex4-1}

\usepackage{amsmath,amsthm,latexsym,amssymb,amsfonts,epsfig}

\usepackage[utf8]{inputenc}
\usepackage{graphicx, texdraw}
\usepackage{bm}

\usepackage[normalem]{ulem} 

\usepackage{type1ec}
\usepackage{mathtools}
\usepackage{mathrsfs}

\usepackage[outer=1.34cm, inner = 1.34cm, top= 2.2cm, bottom = 2.2cm]{geometry}

\usepackage{enumerate}

\usepackage[colorlinks=true,linkcolor=MidnightBlue,urlcolor=MidnightBlue,citecolor=MidnightBlue,anchorcolor=MidnightBlue]{hyperref}

\usepackage{csquotes}

\usepackage{color}
\usepackage[dvipsnames]{xcolor}

\usepackage{amsmath}
\usepackage{amsfonts}
\usepackage{amssymb}

\input{commands}

\usepackage{ulem}

\usepackage{setspace}

\begin{document}

\title{Dynamics of a microswimmer--microplatelet composite}

\author{Abdallah Daddi-Moussa-Ider$^1$}
\email{ider@thphy.uni-duesseldorf.de}
\affiliation
{$^1$ Institut f\"{u}r Theoretische Physik II: Weiche Materie, Heinrich-Heine-Universit\"{a}t D\"{u}sseldorf, Universit\"{a}tsstra\ss e 1, 40225 D\"{u}sseldorf, Germany\\
$^2$Institute of Theoretical Physics, Faculty of Physics, University of Warsaw, Pasteura 5, 02-093 Warsaw, Poland }

\author{Maciej Lisicki$^2$}
\affiliation
{$^1$ Institut f\"{u}r Theoretische Physik II: Weiche Materie, Heinrich-Heine-Universit\"{a}t D\"{u}sseldorf, Universit\"{a}tsstra\ss e 1, 40225 D\"{u}sseldorf, Germany\\
$^2$Institute of Theoretical Physics, Faculty of Physics, University of Warsaw, Pasteura 5, 02-093 Warsaw, Poland }

\author{Hartmut L\"{o}wen$^1$}
\affiliation
{$^1$ Institut f\"{u}r Theoretische Physik II: Weiche Materie, Heinrich-Heine-Universit\"{a}t D\"{u}sseldorf, Universit\"{a}tsstra\ss e 1, 40225 D\"{u}sseldorf, Germany\\
$^2$Institute of Theoretical Physics, Faculty of Physics, University of Warsaw, Pasteura 5, 02-093 Warsaw, Poland }

\author{Andreas M. Menzel$^1$}
\email{menzel@thphy.uni-duesseldorf.de}
\affiliation
{$^1$ Institut f\"{u}r Theoretische Physik II: Weiche Materie, Heinrich-Heine-Universit\"{a}t D\"{u}sseldorf, Universit\"{a}tsstra\ss e 1, 40225 D\"{u}sseldorf, Germany\\
$^2$Institute of Theoretical Physics, Faculty of Physics, University of Warsaw, Pasteura 5, 02-093 Warsaw, Poland }

\date{\today}

\begin{abstract}
Guiding active microswimmers by external fields to requested target locations is a promising strategy to realize complex transport on the microscale. To this end, one possibility consists of attaching the microswimmers to orientable passive components. 
Accordingly, we analyze theoretically, using a minimal model, the dynamics of a microswimmer when rigidly attached tangentially to a (significantly larger) microplatelet, here represented by a thin circular disk. 
On this way, we determine the flow field in the whole space induced by a Stokeslet that is located above the center of a spatially fixed rigid disk of no-slip surface conditions. 
Finally, we determine and analyze possible trajectories of the overall composite.
To this end,  the platelet is additionally endowed with a permanent magnetic moment, which allows to steer the motion of the whole composite by a homogeneous external magnetic field. 
As previous experimental studies suggest, related setups may be helpful to guide sperm cells to requested targets or for the purpose of coordinated drug delivery. 
\end{abstract}
\maketitle

\section{Introduction}

The world of slow viscous flows, characterized by the lack of inertia, or, equivalently, by a vanishingly small Reynolds number~\cite{happel12}, encompasses multiple time and length scales, spanning from geophysical creeping flows such as in marine ice sheets~\cite{rallabandi17prl, rallabandi17ice} to the motion of nanoparticles in aqueous environments~\cite{kim13} and at interfaces~\cite{lang}, which are relevant to technological applications in microfluidic devices~\cite{Squires2005}. Creeping flows are also an essential component of microscale active matter and self-propelling particles. There, the resulting biological fluid flows are directly related to the motility of microorganisms~\cite{lauga2016ARFM,lauga09,elgeti15}.

Considerable attention has been given both to the modeling of swimmers in the context of biological locomotion, and to the artificial biomimetic systems capable of self-propulsion by exploiting various types of available fuel~\cite{bechinger16,zottl16,menzel15,marchetti13,illien17}. Some examples include the use of chemically active surfaces for catalyzed decomposition of solutions, which produces local diffusiophoretic surface flows~\cite{anderson89,golestanian05,golestanian07}, local targeted heating inducing thermophoretic motion in the fluid~\cite{Jiang2010} or controlled decomposition of the surrounding medium~\cite{Wurger2010,wurger15}, as well as temperature changes coupled to the elastic properties of model swimmers~\cite{Zhang2019}. 
Other propulsion mechanisms are based on bead-model designs capable of net swimming when the mutual distance between their constituents are periodically changed in a controlled way such that time reversibility is broken~\cite{najafi04, dreyfus05, golestanian08, grosjean16, daddi18,daddi18jpcm, sukhov19, ziegler19, nasouri19}. 
A characteristic feature of these objects is the force- and torque-free swimming of the individual units, which is a low-Reynolds-number flow property generally shared by motile microswimmers. It results from the overall balance between propulsive forces generated by the swimmer and the fluid drag, which leads to the adjustment of the swimming velocity~\cite{taylor51}.

The specific propulsion mechanisms greatly depend on the swimmer geometry. Bacterial cells often rely on the translation--rotation coupling of their helical filaments, which produce thrust force when the bacterial motors induce their rotation~\cite{lauga2016ARFM}. Ciliated eukaryotic cells generate surface flows by a coordinated motion of cilia covering their bodies, leading effectively to a surface flow, which can be represented mathematically using the squirmer model~\cite{blake71cilia}. However, the elucidation of a specific mechanism for a given organism requires the inclusion of all its geometric and kinematic features~\cite{lauga09}.

One area of prospectively exploiting the self-propulsion of such active micromachines contains the context of directed transport on the microscale. For this purpose, artificial self-propelled magnetic colloidal Janus particles were generated that could be steered by external magnetic fields~\cite{baraban12, sprenger19}. 
Similar guidance was achieved for magnetic bubble-driven tubes~\cite{solovev10}. 

Yet, for many applications of microscale transport, biocompatibility is a central issue. Therefore, biological microswimmers may, for instance, prove as vehicles of choice for such promising tasks as directed drug delivery. In this context, sperm cells as driving motors were loaded with drugs and combined with magnetically addressable microscopic steering components~\cite{xu17}. 
Naturally, if only the sperm cells themselves shall be guided to requested target areas, similar strategies of forming microscopic composites by anchoring them to magnetic steering units have proven successful~\cite{magdanz16}.

In this spirit, we study in the present contribution a motile composite consisting of a microswimmer (such as a bacterium or a sperm cell) attached to a (significantly larger) microplatelet. 
We consider this microplatelet to be given by an infinitely thin rigid disk. The microswimmer is attached laterally on the top surface of the disk. For simplicity, we represent its active forcing (resulting, e.g., from the rotation of a flagella bundle or beating of cilia) by a Stokeslet located at a fixed distance above the center of the disk and oriented parallel to the disk surface.
The adsorption of the microswimmer to the rigid microplatelet will effectively and qualitatively change its swimming trajectory. Instead of a tendency of straight propulsion of the single microswimmer, now circling results for the composite system.
Moreover, the disk now acts as a steering unit. We assume it to carry a permanent in-plane magnetic dipole moment. Through a homogeneous external magnetic field, the composite can thus be guided from outside. Depending on the strength of the resulting magnetic torques and the orientation of the magnetic moment relative to the swimming direction, different types of trajectories are found. 

We base our study on analytical solutions to the problem of Stokes flow due to a point force in the presence of a planar no-slip disk of negligible thickness~\cite{kim83,miyazaki84}.
In this context, we include in the Appendix an analytical approach to find the corresponding flow field when the disk is kept rigidly fixed in space, which complements existing routes to this subject. The problem is formulated as a mixed boundary-value problem for finding three harmonic functions that describe the reflected flow due to the presence of the disk. The resulting dual integral equations are solved by a method described by Kim~\cite{kim83} and introduced by Sneddon~\cite{sneddon60} and Copson~\cite{copson61}. We ultimately use the flow solutions to analyze the motion of our microswimmer--microplatelet composite, also in the presence of an external field, which can couple to the dynamics via a permanent magnetic moment of the platelet.

The paper is structured as follows. Firstly, we introduce our model microswimmer--microplatelet composite in Sec.~\ref{mathmodel}, where we discuss in detail its geometry and the symmetry features. Then, in Sec.~\ref{asymmstokes} we describe the solution technique for a point force rigidly attached to the axis of the disk at a fixed height above the disk and pointing parallel to the disk surface. The obtained analytical solutions are used to calculate the resulting hydrodynamic force and torque on the disk due to the presence of the attached microswimmer. We then use these results in Sec.~\ref{trajectories} to elucidate the swimming trajectories of the composite system under different magnitudes and relative orientations of the permanent magnetic moment of the disk and the external magnetic field. We conclude the paper in Sec.~\ref{conclusion}. The mathematical details of deriving the solution to the flow problem described in Sec.~\ref{asymmstokes} are shifted to the Appendix.

\section{\label{mathmodel} Mathematical model}

\begin{figure}
	\centering
	\includegraphics[width=\columnwidth]{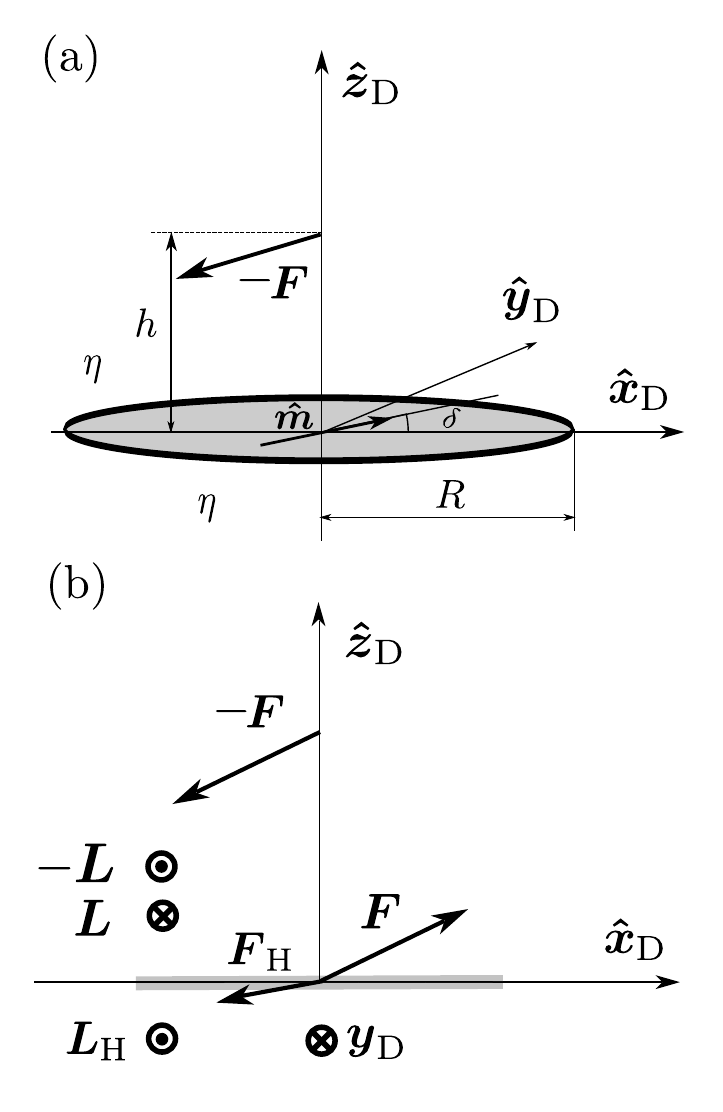}
	\caption{$(a)$~Perspective and ~$(b)$~cross-sectional views of the considered idealized system, representing a microswimmer--microplatelet composite.
	The action of the microswimmer that is attached or adsorbed to the disk is represented by a Stokeslet acting on the fluid at a fixed height~$h$ above the center of the disk. It exerts a force $-\bm{F}$ onto the surrounding fluid. This drives the disk forward through the fluid by a force~$\bm{F}$. Along the same lines, a torque $\bm{L}$ rotates the disk against the fluid [where $-\bm{L}=h\bm{\hat{z}}_\mathrm{D}\times(-\bm{F})$]. 
	The whole space is filled with an otherwise quiescent Newtonian viscous fluid of constant viscosity~$\eta$. 
	Since the fluid is set into motion by the force $-\bm{F}$, an additional hydrodynamic force $\bm{F}_\mathrm{H}$ and torque $\bm{L}_\mathrm{H}$ are acting on the disk. Moreover, the disk carries a permanent magnetic moment $\bm{m}$ aligned within the plane of the disk relatively to $\bm{F}$ by an angle $\delta$. The coordinate frame $\{\bm{\hat{x}}_\mathrm{D},\bm{\hat{y}}_\mathrm{D},\bm{\hat{z}}_\mathrm{D}\}$ represents the frame of reference of the disk. 
	}
	\label{sketch}
\end{figure}

We consider a composite of a microswimmer, for instance, a self-propelled sperm or bacterial cell, attached or adsorbed to a microplatelet. The form of the microplatelet is an ideally thin, rigid, microscopic disk of radius $R$, see Fig.~\ref{sketch}. For our calculations, we assign an orthogonal base unit frame of reference $\{\bm{\hat{x}}_\mathrm{D},\bm{\hat{y}}_\mathrm{D},\bm{\hat{z}}_\mathrm{D}\}$ to the disk, where $\bm{\hat{z}}_\mathrm{D}$ points along the normal of the disk and the origin of this frame of reference coincides with the center of the disk. 
To decipher a mathematical approach to the dynamics of the composite, we are guided by the following reasoning. A microswimmer that is not yet adsorbed or attached to another component but freely propelling through the fluid is force- and torque-free. For instance, considering a bacterial or sperm cell, their rotating flagella bundle or beating cilia work on the surrounding fluid and push it backward. We refer to this net active forcing by the microswimmer on the fluid to achieve self-propulsion by an active force~$-\bm{F}$ acting on the fluid. Through this action, the cell body or head of the microswimmer is pushed forward through the fluid. Friction between the swimmer body and the surrounding fluid takes the fluid along and thus pushes it forward as well. In a simple model and under low-Reynolds-number conditions, this forward forcing of the fluid through friction can be represented by a force~$+\bm{F}$. Together, the two forces $-\bm{F}$ and $+\bm{F}$ form a force dipole acting on the fluid with vanishing total force~\cite{fily12, baskaran09, menzel16, heidenreich16,adhyapak17, adhyapak18, reinken18, daddi18nematic}. 

In our situation, the microswimmer is rigidly attached, adsorbed, or anchored to the (significantly larger) disk. Still, its active propulsion force $-\bm{F}$ on the fluid persists. It now drives the whole composite forward. However, since the disk is much more bulky than the small head or cell body of the microswimmer, we may now neglect the frictional force of the head or cell body of the microswimmer with the fluid. The main frictional contribution now arises because the disk is pushed through the fluid. We therefore may assign the corresponding frictional force $+\bm{F}$ dragging the fluid forward to the disk, instead of to the head or cell body of the microswimmer. 

Our mathematical representation of this situation is illustrated in Fig.~\ref{sketch}. We represent the active backward forcing of the fluid by a Stokeslet $-\bm{F}=-F\bm{\hat{x}}_\mathrm{D}$ oriented parallel to the surface of the disk and at a fixed height $h$ above the center of the disk, at position $\bm{r}_\mathrm{D}=h\bm{\hat{z}}_\mathrm{D}$. Along the above lines, the forward forcing, by which the disk is thus pushed through the fluid by the mechanical attachment of the microswimmer, is thus $+\bm{F}=F\bm{\hat{x}}_\mathrm{D}$. Except for the action of the composite, the surrounding fluid is quiescent and of viscosity $\eta$. 

Furthermore, from the location of the Stokeslet at finite distance above the disk, a net torque $-\bm{L}=\bm{r}_\mathrm{D}\times(-\bm{F})$ is exerted on the fluid by the active forcing of the fluid through the Stokeslet $-\bm{F}$. Through the rigid attachment of the microswimmer or Stokeslet to the disk, the corresponding countertorque $+\bm{L}$ is mechanically transmitted to the disk. By this torque $+\bm{L}$, the disk is rotated, against the frictional resistance of the fluid. (Together, the two torques $-\bm{L}$ and $+\bm{L}$ lead to a vanishing total torque on the surrounding fluid, in agreement with the corresponding low-Reynolds-number condition of torque-free-swimming of the whole composite). In our geometry, depicted in Fig.~\ref{sketch}, the torque $+\bm{L}$ is pointing into the direction $\bm{\hat{y}}_\mathrm{D}$. Also for the rotational dynamics of the whole composite, we neglect the friction of the cell body or head of the microswimmer with the surrounding fluid against the frictional drag between the fluid and the significantly larger disk. 

In addition to that, the microplatelet is endowed with a permanent magnetic moment $\bm{m}$ rigidly anchored to the disk frame and pointing into a fixed direction within the plane of the disk. In the presence of a homogeneous, constant, external magnetic field, an aligning magnetic torque thus results on the disk, which allows to guide the whole composite and thus the adsorbed microswimmer, for instance, a sperm cell, into a requested direction. The angle of $\bm{m}$ relative to the axis $\bm{\hat{x}}_\mathrm{D}$ is denoted as $\delta$, so that
\begin{equation}
	\vect{m} = m \left( \cos\delta \, \vect{\hat{x}}_\mathrm{D} + \sin\delta \, \vect{\hat{y}}_\mathrm{D} \right) \, .
\end{equation}

We denote by $\left( \vect{\hat{x}}_0, \vect{\hat{y}}_0, \vect{\hat{z}}_0  \right)$ the basis unit vectors associated with the laboratory frame. Without loss of generality, we set the external magnetic field {in the fixed laboratory frame of reference} as $\vect{B}= B\vect{\hat{x}}_0$.

There is still more to that. Since the Stokeslet sets the surrounding fluid into motion, corresponding fluid flows act on the disk as well. Through the viscous stresses on the surface of the disk, additional hydrodynamic forces and torques result that we denote as $\bm{F}_\mathrm{H}$ and $\bm{L}_\mathrm{H}$, respectively. They will be calculated explicitly below. For the moment, collecting all these forces and torques on the disk, we can, because of the linearity of the underlying Stokes equations, formulate the resulting overall velocity $\bm{V}$ and angular velocity $\bm{\Omega}$ as
\begin{subequations}\label{SwimmingVelo-Init-Eqns}
	\begin{align}
		\vect{V} &= \boldsymbol{\mu}^{tt} \left( \vect{F} + \vect{F}_\mathrm{H} \right) , \\
		\boldsymbol{\Omega} &= \boldsymbol{\mu}^{rr} \left( \vect{r}_\mathrm{D} \times \vect{F}   + \vect{m} \times \vect{B} + \vect{L}_\mathrm{H}  \right). 
	\end{align}
\end{subequations}

Here, the components of the translational and rotational hydrodynamic self-mobility tensors $\bm{\mu}^{tt}$ and $\bm{\mu}^{rr}$ of the disk expressed in the body-fixed frame of reference read~\cite{oberbeck76, sampson91, ray36, roscoe49, gupta57, tanzosh96, sherwood12}
\begin{subequations}
	\begin{align}
		\mu_{ij}^{tt} &= \mu_\parallel^{tt} \left( \delta_{ij} - \delta_{i3}\delta_{j3} \right) + \mu_\perp^{tt} \delta_{i3} \delta_{j3} \, , \\
		\mu_{ij}^{rr} &= \mu^{rr} \delta_{ij} \, , 
	\end{align}
\end{subequations}
with the mobility coefficients for a negligibly thin disk being
\begin{equation}
	\mu_\parallel^{tt} = \frac{3}{32\eta R} \, , \qquad
	\mu_\perp^{tt} = \frac{1}{16\eta R} \, , \qquad
	\mu^{rr} = \frac{3}{32\eta R^3} \, .
\end{equation}

Notably, the mobility coefficient~$ \mu_\parallel^{tt}$ associated with edgewise translation is larger than that associated with broadside translation along the axis of symmetry.
We note that no coupling between translational and rotational degrees of freedom occurs for this disk immersed in a bulk fluid in the Stokes regime due to the symmetry.

The hydrodynamic force $\bm{F}_\mathrm{H}$ and torque $\bm{L}_\mathrm{H}$ acting on the disk from the fluid flow driven by the force $-\bm{F}$ can be calculated explicitly. For this purpose, we first suppose the whole configuration, that is, the Stokeslet and the disk, to be rigidly fixed in space. We can solve the flow problem under these circumstances. From the fluid flow driven by $-\bm{F}$, viscous stresses result on the surface of the disk, where we assume no-slip conditions. 
From these stresses, we can calculate the corresponding hydrodynamic $\bm{F}_\mathrm{H}$ and torque $\bm{L}_\mathrm{H}$ acting on the disk when it is fixed in space.
In other words, the disk is actually maintained fixed in space by applying some effective counterforce $-\bm{F}_\mathrm{H}$ and countertorque $-\bm{L}_\mathrm{H}$ from outside.
Thus, we know how to return to the situation of the freely movable disk again. We simply have to lift the fixation force $-\bm{F}_\mathrm{H}$ and fixation torque $-\bm{L}_\mathrm{H}$ by a corresponding counterforce $-(-\bm{F}_\mathrm{H})$ and countertorque $-(-\bm{L}_\mathrm{H})$. The latter act on the disk and appear in Eqs.~(\ref{SwimmingVelo-Init-Eqns}).

More precisely, in the situation of keeping the disk fixed, the flow velocity and pressure fields can be obtained by introducing an image system represented by a set of harmonic functions, which can be expanded into Fourier-Bessel integrals. 
Implementing the relevant boundary conditions, the solution for the hydrodynamic flow field can be formulated using a mixed boundary value problem, which can then be transformed into a system of dual integral equations for the expansion coefficients on the inner and outer regions of the domain.
An analogous resolution procedure has previously been utilized by Kim~\cite{kim83} to address the flow problem due to a Stokeslet directed along the normal of the disk.

{
Before we proceed further, it is worth stressing that, in our idealized setup, only the disk experiences the frictional force with the surrounding fluid.
Accordingly, here the hydrodynamic center of mobility (h.c.m.) coincides with the origin of the disk.
In general, for an extended object of finite size attached to the disk, the effective frictional force should be exerted on the h.c.m.\ of the composite.
This aspect is beyond the scope of the present article and merits investigation in a future work.
}

\section{\label{asymmstokes} Stokeslet near a disk: a dual integral equation method}

\subsection{Formulation of the mathematical flow problem}\label{SolFormulation}

Under creeping flow conditions, the fluid dynamics are governed by the incompressible Stokes equations~\cite{happel12,kim13}
\begin{subequations}\label{StokesGleischungen}
	\begin{align}
		-\bNabla p + \eta \bNabla^2 \vect{v} - \vect{F} \delta (\R-\R_\mathrm{D}) &= \vect{0}  \, , \\
		\bNabla \cdot \vect{v} &= 0 \, , 
	\end{align}
\end{subequations}
where $p$ and~$\vect{v}$ denote, respectively, the pressure and fluid velocity fields at position~$\vect{r} = x \,\vect{\hat{x}}_\mathrm{D} + y \,\vect{\hat{y}}_\mathrm{D} + z \,\vect{\hat{z}}_\mathrm{D}$ expressed in the frame of reference attached to the disk, the latter being held fixed in space.

{
The force~$\vect{F}$ can be decomposed into the components along the directions normal and tangential to the disk, respectively denoted by~$\vect{F}_\perp = F_\perp \, \vect{\hat{z}}_\mathrm{D} $ and $\vect{F}_\parallel = F_\parallel \, \vect{\hat{x}}_\mathrm{D}$. 
The solution of the flow problem for an axisymmetric configuration of a Stokeslet acting normal to the surface of a rigid disk at a certain distance above the disk has previously been derived by Kim~\cite{kim83}.
We will thus make use of this result and, using an analogous dual integral equation approach, derive the corresponding solution for a Stokeslet acting tangentially to the disk.
The total flow field can then be obtained by superposition of the two solutions owing to the linearity of the Stokes equations.
}

In an unbounded, infinitely extended fluid (i.e.\ in the absence of the confining disk), the solution of Eqs.~\eqref{StokesGleischungen} {for a Stokeslet of tangential orientation, located at a certain distance above the center of the disk} is expressed in terms of the free-space Green's function, also known as the fundamental solution of Stokes flow.
In cylindrical coordinates~$(r,\phi,z)$, the components of the velocity field due to the point force in an otherwise quiescent fluid are given by
\begin{subequations}
	\begin{align}
		{v_r}_\parallel^\mathrm{S} &= \frac{-F_\parallel}{8\pi\eta s} \left( 1 + \frac{r^2}{s^2} \right) \cos\phi \, , \\
		{v_\phi}_\parallel^\mathrm{S} &= \frac{F_\parallel}{8\pi\eta s} \, \sin\phi \, , \\
		{v_z}_\parallel^\mathrm{S} &= \frac{-F_\parallel}{8\pi\eta} \frac{r \left(z-h\right)}{s^3} \, \cos\phi \, , 
	\end{align}
\end{subequations}
wherein~$s = |\R-\R_\mathrm{D}| = \left(r^2 + (z-h)^2\right)^{1/2}$ stands for the distance from the Stokeslet position.
The corresponding solution for the pressure field reads
\begin{equation}
	p_\parallel^\mathrm{S} = -\frac{F_\parallel r}{4\pi s^3} \, \cos\phi \, .
\end{equation}

Thanks to the linearity of the Stokes equations, the solution of the flow problem near a rigid disk of no-slip surface conditions can conveniently be written as a superposition of the solution in an infinitely extended fluid and a complementary solution that is needed to satisfy the regularity and boundary conditions~\cite{daddi16c, daddi18coupling}.
Then,
\begin{equation}
	\vect{v} = \vect{v}_\parallel^\mathrm{S} + \vect{v}_\parallel^* \, , \qquad
	p_\parallel = p_\parallel^\mathrm{S} + p_\parallel^* \, ,  
\end{equation}
where~$\vect{v}_\parallel^*$ and~$p_\parallel^*$ denote the complementary solution, also commonly referred to as the image solution.

For an asymmetric Stokes flow, the solution of the homogeneous equations of fluid motion can be expressed in terms of three harmonic functions $\Pi$, $\Psi$, and $\Omega$, as~\cite{shail87}
\begin{subequations}
	\begin{align}
		\vect{v}_\parallel^* &= \bNabla  \Pi + z \bNabla \left( \Psi + \frac{\partial \Pi}{\partial z} \right) 
		- \left( \Psi + \frac{\partial \Pi}{\partial z} \right) \vect{\hat{z}}_\mathrm{D}
		+ \bNabla \times \left( \Omega \, \vect{\hat{z}}_\mathrm{D} \right) \, , \notag \\
		p_\parallel^* &= 2\eta \, \frac{\partial}{\partial z} \left( \Psi + \frac{\partial \Pi}{\partial z} \right) \, , \notag
	\end{align}
\end{subequations}
with
\begin{equation}
	\bNabla^2 \Pi = \bNabla^2 \Psi = \bNabla^2 \Omega = 0\, ,  \label{HarmonikEqs}
\end{equation}
where the gradient~$\bNabla$ and Laplace $\bNabla^2$ operator of a scalar quantity~$f$ are, respectively,  given in the cylindrical coordinates system by
\begin{subequations}
	\begin{align}
		\bNabla f &= \frac{\partial f}{\partial r} \, \vect{\hat{x}}_\mathrm{D}  + \frac{1}{r} \frac{\partial f}{\partial \phi} \, \vect{\hat{y}}_\mathrm{D} + \frac{\partial f}{\partial z} \, \vect{\hat{z}}_\mathrm{D} \, , \\ 
		\bNabla^2 f  &= \frac{1}{r} \frac{\partial }{\partial r} \left( r \frac{\partial f }{\partial r} \right)
		+ \frac{1}{r^2} \frac{\partial^2 f}{\partial \phi^2} 
		+ \frac{\partial^2 f}{\partial z^2} \, .
	\end{align}
\end{subequations}

The projected components of the fluid velocity field are given by
\begin{subequations}
	\begin{align}
		{v_r}_\parallel^* &= \frac{\partial}{\partial r} \left( \Pi + z \, \frac{\partial \Pi}{\partial z} \right) + z \, \frac{\partial \Psi}{\partial r} + \frac{1}{r} \frac{\partial \Omega}{\partial \phi} \, , \\
		{v_\phi}_\parallel^* &= \frac{1}{r} \frac{\partial}{\partial \phi} \left( \Pi + z \, \frac{\partial \Pi}{\partial z} \right) + \frac{z}{r} \frac{\partial \Psi}{\partial \phi} - \frac{\partial \Omega}{\partial r} \, , \\
		{v_z}_\parallel^* &= z \, \frac{\partial}{\partial z} \left( \Psi + \frac{\partial \Pi}{\partial z} \right) - \Psi \, .
	\end{align}
\end{subequations}

Since $\Pi$, $\Psi$, and $\Omega$ are harmonic functions, the solution of Eq.~\eqref{HarmonikEqs} can be expressed in terms of infinite series of Fourier-Bessel integrals~\cite{daddi17pof, daddi18acta}.
By requiring the regularity condition that both $\left|\vect{v}_\parallel^*\right|$ and $p_\parallel^*$ must vanish as $r\to \infty$ or~$z \to \pm \infty$, the general solution can be presented as a superposition of Bessel functions with exponential weights and unknown wavelength-dependent coefficients $\pi_k^{\pm} (\lambda)$, $\psi_k^{\pm} (\lambda)$, and~$\omega_k^{\pm} (\lambda)$, for~$k=0,1,\dots$, that will be determined from the boundary conditions in the plane of the disk at $z=0$. The plus sign in the superscripts denotes a quantity in the upper-half space~$z \ge 0$, whereas the minus sign denotes the corresponding quantity in the lower-half space, for which~$z \le 0$. We now write the solution as
\begin{subequations}\label{PiPsiOmega}
	\begin{align}
		\Pi^{\pm} &= \frac{-F_\parallel}{8\pi\eta} \sum_{k\ge 0} \cos (k\phi) 
		      \int_0^\infty \pi_k^{\pm} (\lambda) J_k (\lambda r) e^{-\lambda |z|} \, \Intd \lambda \, , \\
		\Psi^{\pm} &= \frac{-F_\parallel}{8\pi\eta} \sum_{k\ge 0} \cos (k\phi) 
		      \int_0^\infty \psi_k^{\pm} (\lambda) J_k (\lambda r) e^{-\lambda |z|} \, \Intd \lambda \, , \\
		\Omega^{\pm} &= \frac{-F_\parallel}{8\pi\eta} \sum_{k\ge 0} \sin (k\phi) 
			  \int_0^\infty \omega_k^{\pm} (\lambda) J_k (\lambda r) e^{-\lambda |z|} \, \Intd \lambda \, .
	\end{align}
\end{subequations}

\subsection{Boundary conditions and dual integral equations}\label{BCs}

The boundary conditions at $z=0$ represent the continuity of the flow velocity at $z=0$, vanishing velocity on the surface of the disk, and continuity of normal stresses on both sides of the disk, outside the disk, respectively, 
\begin{subequations}
	\begin{align}
		\left. \vect{v}_\parallel^+ - \vect{v}_\parallel^- \right|_{z=0}\,  &= 0 \, , \label{VeloContinuity}  \\
		\left. \vect{v}_\parallel^+ \right|_{z=0} \, = \, \left. \vect{v}_\parallel^- \right|_{z=0} \, &= 0 \,   \quad \text{for~} r<R \, , \label{VanishingVeloInside} \\
		\left. \left( \boldsymbol{\sigma}_\parallel^+ - \boldsymbol{\sigma}_\parallel^- \right) \cdot \vect{\hat{z}}_\mathrm{D} \right|_{z=0} \,  &= 0 \,  \quad \text{for~} r>R  \, , \label{ContinuityStressOutside}
	\end{align}
\end{subequations}
where~$\boldsymbol{\sigma}^\pm \cdot \vect{\hat{z}}_0$ represents the hydrodynamic stress vector on the plane of the disk, with its components given by
\begin{subequations}\label{stresses}
	\begin{align}
		{\sigma_{rz}}_\parallel^\pm &= \eta \left( \frac{\partial {v_r}_\parallel^\pm}{\partial z} + \frac{\partial {v_z}_\parallel^\pm}{\partial r} \right) \, , \label{stressRZ} \\
		{\sigma_{\phi z}}_\parallel^\pm &= \eta \left( \frac{\partial {v_\phi}_\parallel^\pm}{\partial z} + \frac{1}{r} \frac{\partial {v_z}_\parallel^\pm}{\partial \phi}  \right) \, , \label{stressPhiZ} \\
		{\sigma_{zz}}_\parallel^\pm &= -p_\parallel^\pm + 2\eta \, \frac{\partial {v_z}_\parallel^\pm}{\partial z} \, . \label{stressZZ}
	\end{align}
\end{subequations}

To satisfy the natural continuity of the fluid velocity field at~$z=0$, given by Eq.~\eqref{VeloContinuity}, we require~$\pi_k^+ \equiv \pi_k^-$, $\psi_k^+ \equiv \psi_k^-$, and~$\omega_k^+ \equiv \omega_k^-$, for all $k$.
In the remainder of this manuscript, we will thus drop the $\pm$ sign and denote the unknown wavenumber-dependent functions in the upper- and lower-half domains by~$\pi_k(\lambda)$, $\psi_k(\lambda)$, and~$\omega_k(\lambda)$.

Next, applying the no-slip boundary conditions at the surface of the disk, as prescribed by Eq.~\eqref{VanishingVeloInside}, yields the following integral equations for $r<R$
\begin{subequations} \label{InnerProblemWithInfSeries}
	\begin{align}
		\sum_{k\ge 0} \cos (k\phi) 
		\int_0^\infty 
		\Gamma_k^r (r, \lambda) \, \Intd \lambda &= - \frac{\left(2r^2+h^2\right) \cos\phi}{\left(r^2+h^2\right)^{3/2}} \, , \\
		\sum_{k\ge 0} \sin (k\phi) 
		\int_0^\infty \Gamma_k^\phi (r, \lambda) \,  \Intd \lambda &=  \frac{\sin\phi}{\left(r^2+h^2\right)^{1/2}} \, , \\
		\sum_{k\ge 0} \cos (k\phi) 
		\int_0^\infty \Gamma_k^z (r, \lambda)  \, \Intd \lambda &=  \frac{hr \cos\phi}{\left(r^2+h^2\right)^{3/2}} \, ,
	\end{align}
\end{subequations}
where we have defined for convenience the integrands
\begin{subequations}
	\begin{align}
		\Gamma_k^r (r, \lambda)  &= \frac{k}{r} \left( \pi_k(\lambda)+\omega_k(\lambda) \right) J_k(\lambda r)
				-\lambda \pi_k(\lambda) J_{k+1} (\lambda r) \, , \notag \\
		\Gamma_k^\phi (r, \lambda)  &= -\frac{k}{r} \left( \pi_k(\lambda) + \omega_k(\lambda) \right) J_k(\lambda r)
				+ \lambda \omega_k(\lambda) J_{k+1} (\lambda r) \, , \notag \\
		\Gamma_k^z (r, \lambda)  &= -\psi_k(\lambda) J_k (\lambda r) \, . \notag
	\end{align}
\end{subequations}
The right-hand sides of Eqs.~\eqref{InnerProblemWithInfSeries} derive from the Stokeslet contribution to the total flow field,  in addition to the image solution $\vect{v}^*$. 
Above, the infinite sums can be dealt with by using the orthogonality property of Fourier series components to obtain
\begin{subequations}\label{EqsInnerGamma}
	\begin{align}
		\int_0^\infty 
		\Gamma_k^r (r, \lambda) \, \Intd \lambda
		&= - \frac{2r^2+h^2 }{\left(r^2+h^2\right)^{3/2}} \, \delta_{k1}\, , \label{EqVR} \\
		\int_0^\infty 
		\Gamma_k^\phi (r, \lambda)  \, \Intd \lambda 
		&=  \frac{\delta_{k1}}{\left(r^2+h^2\right)^{1/2}} \, , \label{EqVPhi} \\
		\int_0^\infty 
		\Gamma_k^z (r, \lambda) \, \Intd \lambda 
		&= \frac{hr }{\left(r^2+h^2\right)^{3/2}} \, \delta_{k1} \, .
	\end{align}
\end{subequations}
We now make use of the recurrence identity~\cite{abramowitz72}
\begin{equation}
	J_{k-1}(\lambda r) + J_{k+1}(\lambda r) = \frac{2k}{\lambda r} \, J_k (\lambda r) 
	\label{RecurrenceBessel}
\end{equation}
and introduce the following shorthand notations,
\begin{subequations} \label{InnerProblemFinalized2}
	\begin{align}
		f^-_k(r) &= \frac{r^2 }{\left(r^2+h^2\right)^{3/2}} \, \delta_{k1} \, , \label{fkMinus} \\
		f^+_k(r) &= - \frac{\left(3r^2+2h^2\right) }{\left(r^2+h^2\right)^{3/2}} \, \delta_{k1} \, , \label{fkPlus} \\
		f_k(r)   &= - \frac{hr }{\left(r^2+h^2\right)^{3/2}} \, \, \delta_{k1}  \label{fk}
	\end{align}
\end{subequations}
to rewrite Eqs.~\eqref{EqsInnerGamma} as
\begin{subequations}\label{InnerProblemFinalized}
	\begin{align}
		\int_0^\infty 
		\lambda \left( \pi_k(\lambda) - \omega_k(\lambda) \right) J_{k+1}(\lambda r) \, \Intd \lambda &=  f^-_k (r) \, , \\
		\int_0^\infty 
		\lambda \left( \pi_k(\lambda) + \omega_k(\lambda) \right) J_{k-1}(\lambda r) \, \Intd \lambda &= f^+_k(r) \, , \\
		\int_0^\infty 
		\psi_k(\lambda) J_k (\lambda r) \, \Intd \lambda &= f_k(r) \, .  
	\end{align}
\end{subequations}

Equations~\eqref{InnerProblemFinalized} constitute the integral equations for the inner region.
To obtain the corresponding equations for the outer region, we require the continuity of the hydrodynamic stress tensor, which is stated by Eq.~\eqref{ContinuityStressOutside}.
Defining
\begin{align}
	\Lambda_k^r (r,\lambda) &= 
			\frac{k\lambda}{r} \left(2\pi_k(\lambda)+\omega_k(\lambda)\right) J_k(\lambda r)
			- 2\lambda^2 \pi_k(\lambda) J_{k+1}(\lambda r) , \notag \\
	\Lambda_k^\phi (r,\lambda) &= 
			\frac{k\lambda}{r} \left( 2\pi_k(\lambda)+\omega_k(\lambda) \right) J_k(\lambda r) -\lambda^2\omega_k(\lambda)J_{k+1}(\lambda r)   , \notag \\
	\Lambda_k^z (r,\lambda) &= \lambda \psi_k(\lambda)J_k(\lambda r) \, ,	\notag
\end{align}
we obtain three equations
\begin{equation} \label{EqsOuterLambda}
		\int_0^\infty \Lambda_k^j (r,\lambda) \, \Intd \lambda = 0 \, , \quad j \in \{r,\phi,z\} \, .
\end{equation}

Rearranging terms and making use of Eq.~\eqref{RecurrenceBessel}, Eqs.~\eqref{EqsOuterLambda} can be recast in a finalized integral form as
\begin{subequations}\label{OuterProblemFinalized}
	\begin{align}
		\int_0^\infty \lambda^2 \left( 2\pi_k(\lambda)-\omega_k(\lambda) \right) J_{k+1} (\lambda r) \, \Intd \lambda &= 0 \, , \\
		\int_0^\infty \lambda^2 \left( 2\pi_k(\lambda)+\omega_k(\lambda) \right) J_{k-1} (\lambda r) \, \Intd \lambda &= 0 \, , \\
		\int_0^\infty \lambda \psi_k(\lambda)J_k(\lambda r) \, \Intd \lambda &= 0 \, .
	\end{align}
\end{subequations}
Equations~\eqref{OuterProblemFinalized} represent the integral equations for the outer region.

As detailed in the Appendix, the solution of the resulting dual integral equations for the unknown wavenumber-dependent quantities can conveniently be presented in an integral form as
\begin{align}
	\pi_k(\lambda) &= \frac{\lambda^{-\halb}}{4} \int_0^R
	\left( \hat{\xi}_k^{-}(t)J_{k+\halb}(\lambda t) + \hat{\xi}_k^{+}(t)J_{k-\frac{3}{2}} (\lambda t) \right) \Intd t \, , \notag \\
	\omega_k(\lambda) &= \frac{\lambda^{-\halb}}{2} \int_0^R
	\left( -\hat{\xi}_k^{-}(t)J_{k+\halb}(\lambda t) + \hat{\xi}_k^{+}(t)J_{k-\frac{3}{2}} (\lambda t) \right) \Intd t \, , \notag \\
	\psi_k(\lambda) &= \lambda^{\halb} \int_0^R \hat{\psi}_k (t) J_{k-\halb} (\lambda t) \, \Intd t \, , \notag
\end{align}
where the functions appearing in the integrands as coefficients of the Bessel functions are explicitly given by
\begin{align}
	\hat{\xi}_k^-(t) &= \left(\frac{2t}{\pi}\right)^\halb \frac{4h t^2}{\left(t^2 + h^2\right)^2} \, \delta_{k1} \, , \notag \\
	\hat{\xi}_k^+(t) &= \frac{2}{h} \left(\frac{2t}{\pi}\right)^\halb \left( \frac{t^2 \left(t^2 - h^2\right)}{\left(t^2 + h^2\right)^2} - \frac{3R^2 + 4h^2}{3 \left( R^2 + h^2 \right)} \right) \delta_{k1} \, , \notag \\
	\hat{\psi}_k (t) &= -\left(\frac{2t}{\pi}\right)^{\halb} 
	\frac{2h^2 t}{\left(t^2+h^2\right)^2} \, \delta_{k1} \, . \notag
\end{align}
Notably, all the series coefficients vanish except for~$k=1$.

Having obtained expressions for the hydrodynamic flow field caused by a point force acting tangent to a finite-sized disk fixed in space, we next employ this solution to obtain expressions for the resulting hydrodynamic force and torque on the disk.

\subsection{Hydrodynamic force and torque}

Through the fluid flows induced by the force~$-\vect{F}$ that the microswimmer exerts on the fluid, a net hydrodynamic force and torque result on the rigid disk.
From the solution described above
for the resulting flow field, we can calculate the hydrodynamic stress vector.
From there, the hydrodynamic force and torque on the disk follow via integration over the surface of the disk.

The total hydrodynamic force exerted by the surrounding fluid on both sides of the disk is obtained by integrating the stress vector over the top and bottom surfaces of the disk~$A=\{0\le r\le R, 0\le\phi \le 2\pi, z=0\}$.
Specifically,
\begin{equation}
	\vect{F}_\mathrm{H} = \int_A \left( \boldsymbol{\sigma}^+ - \boldsymbol{\sigma}^- \right) \cdot \vect{\hat{z}}_\mathrm{D} \, \Intd A \, .
\end{equation}

{
The total hydrodynamic force exerted on the disk can be decomposed into an out-of-plane component ${F_\mathrm{H}}_\perp \equiv \vect{F}_\mathrm{H} \cdot \vect{\hat{z}}_\mathrm{D}$ resulting from the normal part of the Stokeslet~$\vect{F}_\perp$ and an in-plane component ${F_\mathrm{H}}_\parallel \equiv \vect{F}_\mathrm{H} \cdot \vect{\hat{x}}_\mathrm{D}$ resulting from the tangential part of the Stokeslet.}

{
${F_\mathrm{H}}_\perp$ is calculated as
\begin{equation}
	{F_\mathrm{H}}_\perp = 2 \int_0^R \left. {\sigma_{zz}}_\perp^+ \right|_{z=0} \, 2\pi r \, \Intd r \, .
\end{equation}
By inserting the expression of the normal component of the hydrodynamic stress vector for an axisymmetric Stokeslet as derived by Kim~\cite{kim83} and evaluating the resulting integral, we find
\begin{equation}
	{F_\mathrm{H}}_\perp = -F_\perp \left( 1 - \frac{2}{\pi} \left( \arctan\xi - \frac{\xi}{1+\xi^2} \right) \right) \, , 
\end{equation}
where we have introduced the dimensionless number~$\xi=h/R$.
}

{In addition to that,} the component~${F_\mathrm{H}}_\parallel$ is given by
\begin{equation}
	{F_\mathrm{H}}_\parallel = 2 \int_0^{2\pi} \int_0^R
	\left. \left( {\sigma_{rz}}_\parallel^+ \cos\phi - {\sigma_{\phi z}}_\parallel^+ \sin\phi \right) \right|_{z=0}
	r \, \Intd r \, \Intd \phi \, . \label{Fx}
\end{equation}

By inserting the expressions of the tangential components of the stress vector as given by Eqs.~\eqref{stressRZ} and~\eqref{stressPhiZ} into Eq.~\eqref{Fx} and performing the double integration over the surface of the disk, we obtain
\begin{equation}
	{F_\mathrm{H}}_\parallel = -\frac{Fh}{3\pi \left(R^2+h^2\right)}
	\int_0^R \mathcal{W}(t) \, \Intd t \, , \label{FH-transient}
\end{equation}
where we have defined
\begin{equation}
	\mathcal{W}(t) = \frac{t^4+\left(9R^2+11h^2\right)t^2+h^2\left(3R^2+4h^2\right)}{\left(t^2+h^2\right)^2}  \notag
\end{equation}
and used the fact that (for~$t<R$)
\begin{equation}
	\int_0^\infty J_1(\lambda R) \cos(\lambda t) \, \Intd \lambda = \frac{1}{R} \, . \notag
\end{equation}

After integration of Eq.~\eqref{FH-transient} and rearranging terms, the hydrodynamic force takes the final form
\begin{equation}
	{F_\mathrm{H}}_\parallel = -F_\parallel \left( 1-\frac{2}{\pi} \left( \arctan\xi +\frac{\xi}{3\left(1+\xi^2\right)} \right) \right) \, .
\end{equation}

Due to the action of the asymmetric point force exerted tangent to the surface of the disk, the latter will also be subject to a total hydrodynamic torque given by
\begin{equation}
	\vect{L}_\mathrm{H} = \int_A \left(  \vect{r} \times \left( \boldsymbol{\sigma}^+ - \boldsymbol{\sigma}^- \right) \right) \cdot \vect{\hat{z}}_\mathrm{D} \, \Intd A \, .
\end{equation}

We find that the resulting hydrodynamic torque has only one non-vanishing component $L_\mathrm{H} \equiv \vect{L}_\mathrm{H} \cdot \vect{\hat{y}}_\mathrm{D}$.
{The latter results from the tangential component of the Stokeslet and is given by}
\begin{equation}
	L_\mathrm{H} = 2 \int_0^{2\pi} \int_0^R -r^2 \left. {\sigma_{zz}}_\parallel^+ \right|_{z=0} \cos\phi \, \Intd r \, \Intd \phi \, . \label{Ly}
\end{equation}

By inserting the expression of the normal component of the stress vector stated by Eq.~\eqref{stressZZ} into Eq.~\eqref{Ly} and carrying out double integration over the surface of the disk, we obtain
\begin{equation}
	L_\mathrm{H} = -\frac{4F_\parallel h^2}{\pi} \int_0^R \frac{t^2 \, \Intd t}{\left(t^2+h^2\right)^2}   \, , \label{LH-transient}
\end{equation}
where we have made use of the identity (for~$t<R$)
\begin{equation}
	\int_0^\infty \left( 2\lambda^{-1} J_1(\lambda R)- R J_0(\lambda R) \right) \sin(\lambda t) \, \Intd \lambda = \frac{2t}{R} \, . \notag
\end{equation}

Finally, the total hydrodynamic torque exerted on the disk readily follows upon performing the integration in Eq.~\eqref{LH-transient} as
\begin{equation}
	L_\mathrm{H} = -F_\parallel h \left( 1-\frac{2}{\pi} \left( \frac{\xi}{1+\xi^2} + \arctan\xi\right) \right) \, .
\end{equation}

In particular, for $\xi=0$ (corresponding to a Stokeslet attached to the disk or equivalently to an infinitely extended disk), we obtain~$\vect{F}_\mathrm{H} = -\vect{F}$ and $\vect{L}_\mathrm{H} = \vect{r}_\mathrm{D} \times (-\vect{F})$. In this case, it correctly follows from Eq.~\eqref{SwimmingVelo-Init-Eqns} that~$\vect{V} = \vect{0}$ and $\boldsymbol{\Omega} = \vect{0}$ for a vanishing external magnetic field.


\section{\label{trajectories} Swimming trajectories of the composite}

Having derived the hydrodynamic force and torque exerted on the disk due to the presence of the Stokeslet, we next examine in detail the swimming trajectories performed by the microswimmer--microplatelet composite.
Obtaining for the resultant force and torque exerted on the disk
\begin{subequations}
	\begin{align}
		\vect{F}_\mathrm{D} &= \frac{2F_\parallel}{\pi} \left( \frac{\xi}{3\left(1+\xi^2\right)}+\arctan\xi \right) \vect{\hat{x}}_\mathrm{D} \notag \\
		&{\quad+\, \frac{2 F_\perp}{\pi} \left( \arctan\xi - \frac{\xi}{1+\xi^2} \right) \vect{\hat{z}}_\mathrm{D} } \, , \\
		\vect{L}_\mathrm{D} &= \frac{2F_\parallel h}{\pi} \left( \frac{\xi}{1+\xi^2} + \arctan\xi \right) \vect{\hat{y}}_\mathrm{D} \, ,
	\end{align}
\end{subequations}
the translational and rotational swimming velocities can be written as
\begin{subequations}
	\begin{align}
		\vect{V} &= \boldsymbol{\mu}^{tt} \cdot \vect{F}_\mathrm{D} \, , \\
		\boldsymbol{\Omega} &= \mu^{rr} \left( \vect{L}_\mathrm{D} + \vect{m} \times \vect{B} \right) \, . \label{OmegaTorqueEq}
	\end{align}
\end{subequations}

\begin{figure}
	\centering
	\includegraphics[width=\columnwidth]{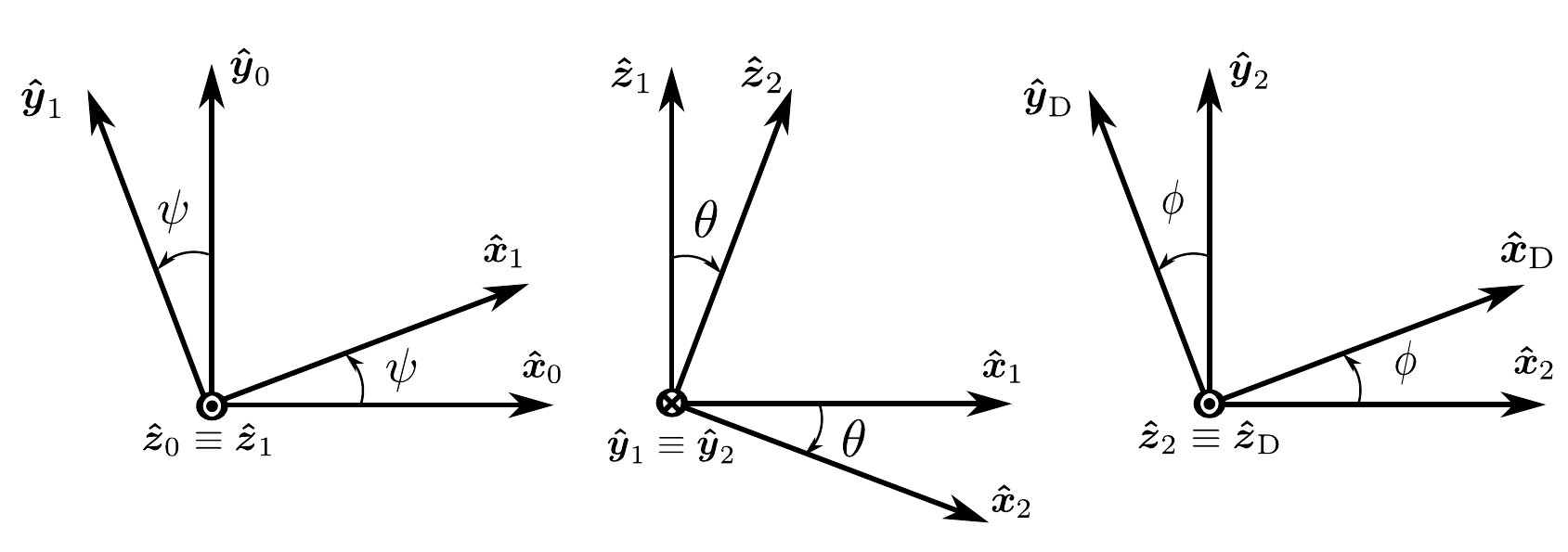}
	\caption{Schematic diagram of Euler angles and axes. 
	The reference frame attached to the disk $\{\hat{\vect{x}}_\mathrm{D}, \hat{\vect{y}}_\mathrm{D}, \hat{\vect{z}}_\mathrm{D}\}$ is obtained via three successive rotations by angles $(\psi, \theta, \phi)$, which represent the precession, nutation, and spin, from the laboratory frame $\{\hat{\vect{x}}_0, \hat{\vect{y}}_0, \hat{\vect{z}}_0 \}$.
	}
	\label{references}
\end{figure}

The position of the moving frame attached to the disk $(\vect{\hat{x}}_\mathrm{D}, \vect{\hat{y}}_\mathrm{D}, \vect{\hat{z}}_\mathrm{D})$ relative to the laboratory frame $(\vect{\hat{x}}_0, \vect{\hat{y}}_0, \vect{\hat{z}}_0)$ can conveniently be described by three Euler angles $(\psi, \theta, \phi)$, commonly denominated as precession, nutation, and proper rotation or spin. The Euler angles are sketched in Fig.~\ref{references}. Within this framework, the angular velocity of the composite is obtained as
\begin{equation}
	\boldsymbol{\Omega} = \dot{\psi} \vect{\hat{z}}_0 + \dot{\theta} \vect{\hat{y}}_1
	+ \dot{\phi} \vect{\hat{z}}_2 \, . \label{OmegaEq}
\end{equation}

The equations governing the translational degree of freedom of the composite in the laboratory frame then read
\begin{equation}
	\begin{split}
		\frac{\Intd \vect{r}}{\Intd t} \equiv \vect{V} &= \mu_\parallel^{tt} {F_\mathrm{D}}_\parallel
		\begin{pmatrix}
			\cos\psi \cos\theta \cos\phi - \sin\psi\sin\phi \\
			\sin\psi\cos\theta\cos\phi + \cos\psi\sin\phi \\
			- \sin\theta \cos\phi
		\end{pmatrix} \\
		&\quad{+ \, \mu_\perp^{tt} {F_\mathrm{D}}_\perp
		\begin{pmatrix}
			\cos\psi\sin\theta \\
			\sin\psi\sin\theta \\
			\cos\theta
		\end{pmatrix} },
	\end{split}
\end{equation}
{where we have defined ${F_\mathrm{D}}_\parallel = \vect{F}_\mathrm{D} \cdot \hat{\vect{x}}_\mathrm{D}$ and ${F_\mathrm{D}}_\perp = \vect{F}_\mathrm{D} \cdot \hat{\vect{z}}_\mathrm{D}$.}

\begin{figure}
	\begin{center}
		\includegraphics[scale=1]{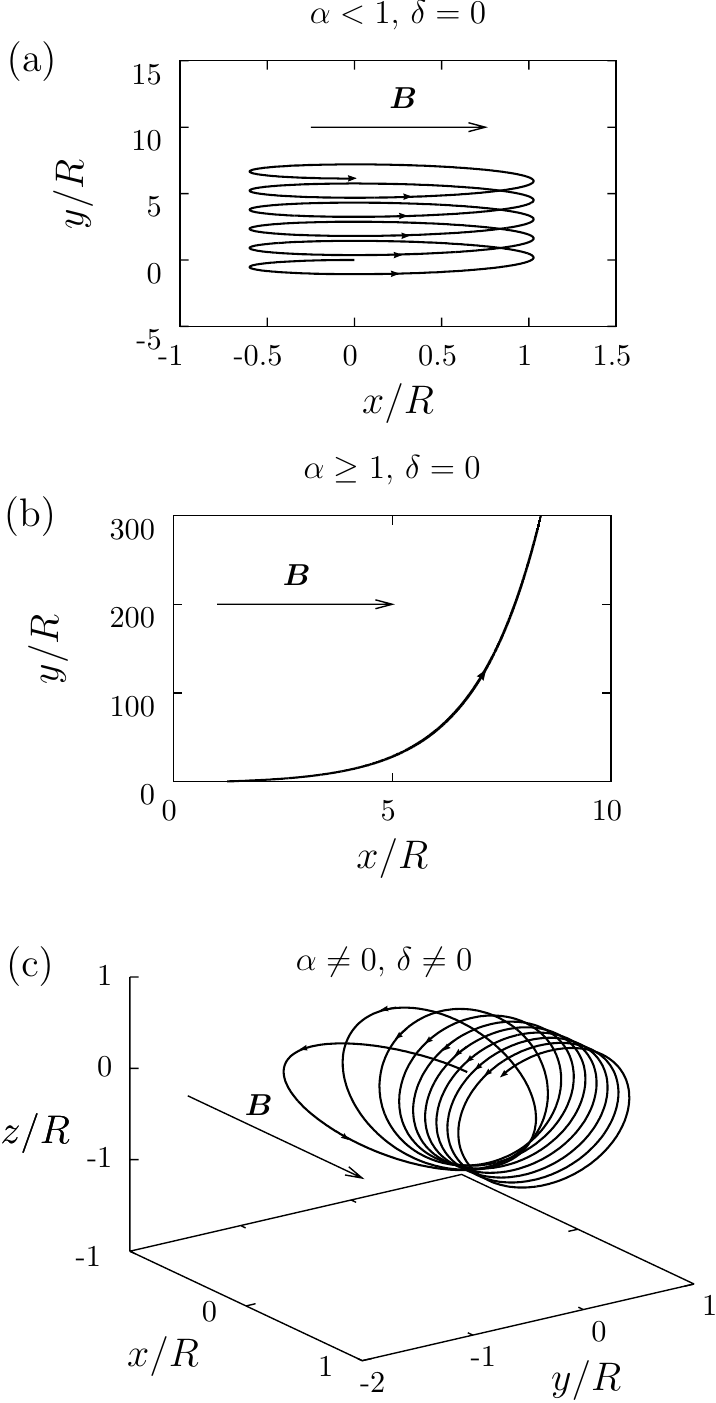}
		\caption{Exemplary swimming trajectories of composites starting from the origin of the laboratory coordinate system for~$(a)$~$\alpha=0.5$ and~$\delta=0$,
		$(b)$~$\alpha=1$ and~$\delta=0$, and~$(c)$~$\alpha=0.2$ and~$\delta=\pi/4$.
		Here, we set the initial angles $\psi_0=\theta_0=\varphi_0=\pi/2$, implying that the trajectories in~$(a)$ and~$(b)$ are confined to the plane~$z_0=0$.
		Moreover, we set $\xi = 1$ {and $\vartheta = 0$.}
		Arrows show the direction of time evolution.
		}
		\label{Traj}
	\end{center}
\end{figure}

\begin{figure}
	\begin{center}
		\includegraphics[scale=1]{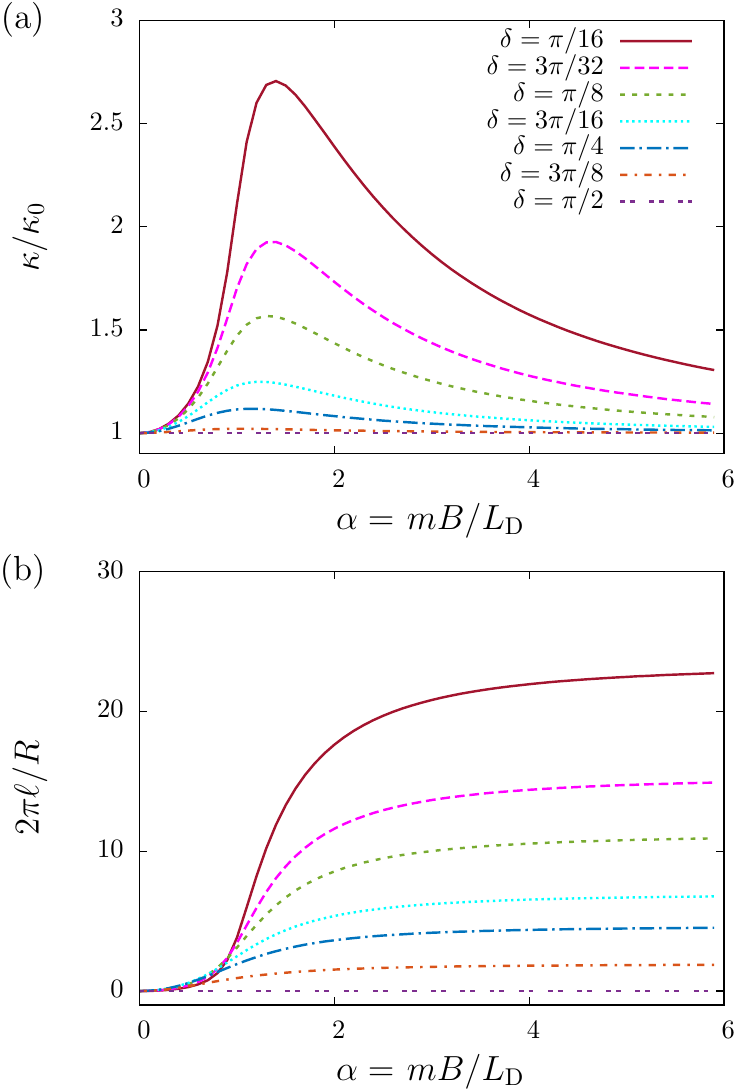}
		\caption{(Color online) 
		$(a)$~Scaled curvature and $(b)$~pitch of the helical swimming trajectory in the steady state versus the scaled strength of the magnetic field for various values of the angle of the magnetic moment relative to the swimming direction.
		Here, we set $\xi=1$ {and $\vartheta = 0$}.
		}
		\label{Rad-Und-Pitch}
	\end{center}
\end{figure}

Combining Eqs.~\eqref{OmegaTorqueEq} and~\eqref{OmegaEq}, and recalling that~$\vect{B} = B \vect{\hat{x}}_0 $, it follows that the temporal variations of the Euler angles are governed by
\begin{equation}
	\begin{pmatrix}
		\dot{\psi} \\
		\dot{\theta} \\
		\dot{\phi} 
	\end{pmatrix}
	=
	\mu^{rr} L_\mathrm{D}
	\begin{pmatrix}
		 \sin\phi \csc\theta - \alpha \cos\psi \sin(\phi+\delta) \\
		 \cos\phi - \alpha \cos\psi\sin\theta \cos(\phi+\delta) \\
		 -\sin\phi\cot\theta -\alpha \sin\psi \cos(\phi+\delta) 
	\end{pmatrix} ,
\end{equation}
where we have defined the dimensionless number
\begin{equation}
	\alpha = \frac{mB}{L_\mathrm{D}} \, ,
\end{equation}
which can be viewed as a measure for the possible magnitude of the torque resulting from the external magnetic field relative to the strength of the hydrodynamic torque acting on the disk due to the fluid flows induced by the Stokeslet.

In order to characterize the swimming trajectories in detail, we decompose the angular velocity into a component~{$\boldsymbol{\Omega}_\parallel = \Omega_\parallel \vect{\hat{t}}$} parallel to the swimming direction and a component~$\boldsymbol{\Omega}_\perp$ perpendicular to it, so that~$\boldsymbol{\Omega} = \boldsymbol{\Omega} _\parallel + \boldsymbol{\Omega}_\perp$.
{Here, we have defined the swimming direction of the composite by the unit vector
\begin{equation}
	\hat{\vect{t}} = \cos\vartheta \, \hat{\vect{x}}_\mathrm{D} + \sin\vartheta \, \hat{\vect{z}}_\mathrm{D} \, , 
\end{equation}
so that
\begin{equation}
	\tan\vartheta = \frac{\mu_\perp^{tt} {F_\mathrm{D}}_\perp}{\mu_\parallel^{tt} {F_\mathrm{D}}_\parallel} 
	= \frac{2}{3} \frac{{F_\mathrm{D}}_\perp}{{F_\mathrm{D}}_\parallel} \, .
\end{equation}}

In terms of Euler angles, {the components of the projected angular velocity are obtained as
\begin{subequations}
	\begin{align}
		\Omega_\parallel &= \alpha \mu^{rr} L_\mathrm{D}  \big( \Lambda_1 \cos\vartheta - \Lambda_2 \sin\vartheta \big) \, , \\
		\boldsymbol{\Omega}_\perp &= \mu^{rr}L_\mathrm{D} \Big(
		\alpha \big( \Lambda_1 \sin\vartheta + \Lambda_2 \cos\vartheta  \big) \left( \sin\vartheta \, \vect{\hat{x}}_\mathrm{D} - \cos\vartheta \, \vect{\hat{z}}_\mathrm{D} \right) \notag \\
		&\quad+\big( 1 - \alpha \cos\delta \cos\psi\sin\theta \big) \vect{\hat{y}}_\mathrm{D}  \Big) \, ,
	\end{align}
\end{subequations}
where we have defined
\begin{subequations}
	\begin{align}
		\Lambda_1 &= \sin\delta \cos\psi\sin\theta \, , \\
		\Lambda_2 &= \sin\psi\cos(\phi+\delta) + \cos\psi\cos\theta \sin(\phi+\delta) \, .
	\end{align}
\end{subequations}}

We now determine the instant curvature and unsigned torsion of the trajectory from the translational and rotational swimming velocities as~\cite{crenshaw93, friedrich09}
\begin{subequations}
	\begin{align}
		\kappa &= \frac{\left| \boldsymbol{\Omega} \times \vect{V} \right|}{\left|\vect{V}\right|^2} = \frac{\left| \boldsymbol{\Omega}_\perp \right|}{\left| \vect{V} \right|} \, , \\
		\tau &= \,\frac{\left| \boldsymbol{\Omega} \cdot \vect{V} \right|}{\left|\vect{V}\right|^2} \, =
		\frac{\left|\Omega_\parallel \right|}{\left| \vect{V} \right|} \, .
	\end{align}
\end{subequations}

In the absence of an external magnetic field $(\alpha=0)$, it follows that~$\tau = 0$.
In this case, the composite will swim along a circular path with a constant curvature
{
\begin{equation}
 \kappa_0 = \left(1+\tan^2\vartheta\right)^{-\halb} \frac{\xi}{R} 
 \frac{\frac{\xi}{1+\xi^2} + \arctan\xi}{\frac{\xi}{3\left(1+\xi^2\right)}+\arctan\xi} \, .	  	
\end{equation}}
{
Accordingly, the curvatures amount to the maximum value when $\vartheta = 0$, for which the Stokeslet is directed parallel to the surface of the disk.
For $\vartheta = \pi/2$, the Stokeslet is directed normal to the surface of the disk and it follows that $\kappa = 0$, thus the composite swims along a straight trajectory without changing its orientation.}

{
Apart from that, we do not observe $\vartheta$ to change qualitatively the resulting swimming trajectories.
Therefore, we now set $\vartheta = 0$.
}

A few exemplary trajectories of microswimmer--microplatelet composites initially starting from the origin of the laboratory frame of reference are presented in Fig.~\ref{Traj}. 
For~$\alpha \ne 0$ and $\delta=0$, implying that~$\vect{m} \parallel \vect{\hat{x}}_\mathrm{D}$, it readily follows that the torsion~$\tau=0$.
Thus, the composite moves in the plane set by the initial orientation, as shown in Fig.~\ref{Traj}$(a,b)$.
From our numerical evaluation for different sets of parameter values, we made the following observations.
When~$\alpha < 1$, then the composite moves along a cycloidal path~\cite{lawrence13} in a plane normal to~$\boldsymbol{\Omega}_\perp$.
This behavior is analogous to that of a noise-free circle swimmer moving under the action of a constant external force~\cite{lowen19}.
In contrast to that, when~$\alpha \ge 1$, then both~$\kappa$ and~$\tau$ in the steady state vanish.
Correspondingly, the composite after a transient evolution will move along a straight trajectory.
For~$\alpha \ne 0$ and $\delta \ne 0$, the swimming trajectories in the steady state, after a transient initial regime, in our numerical observations were found to form regular helices of radius $r=\kappa/\left(\kappa^2 + \tau^2\right)$, a pitch $2\pi \ell = 2\pi\tau /\left( \kappa^2 + \tau^2 \right)$, and a helix angle $\varepsilon = \arctan \left(\ell/r\right)$, see Fig.~\ref{Traj}$(c)$.

The radius and pitch of the steady-state helical trajectory found numerically after a transient reorientation depend on the relative strength of the magnetic and hydrodynamic forcing, as quantified by $\alpha$. This dependence of these geometric features on the increasing field strength is illustrated in Fig.~\ref{Rad-Und-Pitch}. While the pitch of the helix increases monotonically with increasing field strength, the curvature of the helix depends non-trivially on the magnetic field strength, where the curvature reaches a maximum at some intermediate value of the scaled magnitude of the magnetic field.

In this context, analogous helical trajectories have been observed for anisotropic self-propelled colloidal particles moving under gravity~\cite{tenHagen14, campbell17}, and self-diffusiophoretic active particles anisotropically covered with activity and mobility patches~\cite{lisicki18}.
In addition, it has been demonstrated that biaxial self-propelling active particles with arbitrary shape can swim along helical or even superhelical trajectories~\cite{wittkowski12}.
For biological microswimmers, helical swimming paths have been observed in various types of bacteria such as \textit{Leptospira}~\cite{berg79}, \textit{Spiroplasma}~\cite{yang09}, \textit{Caulobacter crescentus}~\cite{liu14}, and \textit{Helicobacter pylori}~\cite{constantino16}.
These experimentally observed wiggling trajectories are connected to the rotations of their chiral flagella bundles~\cite{hyon12}. 
Meanwhile, sea urchin sperm cells~\cite{crenshaw96, corkidi08, kaupp08, hilfinger08} and human sperms~\cite{su12} have also been shown to swim along well-defined helical paths.
Further studies have accounted for the effect of thermal fluctuations on the helical motion of active Brownian particles~\cite{friedrich08, sevilla16}.

\section{\label{conclusion} Conclusions}

In this work, we have analyzed the dynamics of a microswimmer--microplatelet composite propelling at low Reynolds numbers through an otherwise quiescent, incompressible, and viscous environment. The action of the microswimmer was represented by a Stokeslet pushing the fluid to the back at a fixed height above the disk that represents the microplatelet. We derived the translational and rotational equations of motion for this composite, including hydrodynamic effects. On the way, we also presented an alternative way to analytically determine the fluid flow induced by the Stokeslet when the disk is rigidly fixed in space under no-slip surface conditions. 

In addition, we numerically evaluated the equations of motion of the composite. Without further action, in general, circular trajectories result. We then additionally considered the disk to feature a permanent in-plane magnetic moment, so that it is subject to an additional torque in a homogeneous external magnetic field. With the magnetic moment pointing into the swimming direction, cycloidal or steady-state straight trajectories were found with increasing magnetic field amplitudes. In the more general case of the in-plane magnetic moment not being aligned with the swimming direction, helical trajectories were observed. Our results may be important in the context of recent attempts to guide sperm cells or other microswimmers through the combination with non-active steering components to requested target areas.

{
For some swimming microorganisms, such as flagellated bacteria, which perform a screw-like motion to achieve propulsion~\cite{berg73}, accounting for the self-generated active torque may become important. In the present context, to include this aspect, one would require the knowledge of the flow field generated by a point torque acting near a finite-sized disk under no-slip surface conditions. 
This is an interesting extension of the problem for future studies.
}

\begin{acknowledgments}
	We thank Soudeh Jahanshahi for useful discussions.
	A.D.M.I, H.L., and A.M.M. gratefully acknowledge support from the DFG (Deutsche Forschungsgemeinschaft) through the projects DA~2107/1-1, LO~418/16-3, and ME~3571/2-2.
\end{acknowledgments}

\appendix 

\section{\label{appendix1} Solution of the dual integral equations}

In this Appendix, we present an analytical approach to obtain closed-form solutions for the flow field that we have formulated as a mixed boundary value problem in the main body of the paper. 

\subsection{Solution for~$\psi_k$}

We first consider the dual integral equations
\begin{subequations} \label{DualIntEqsPsi}
	\begin{align}
		\int_0^\infty \psi_k(\lambda)J_k(\lambda r) \, \Intd \lambda &= f_k(r) \,  \qquad\,\, (0<r<R) \, , \label{EqInnerPsi} \\
		\int_0^\infty \lambda \psi_k(\lambda) J_k(\lambda r) \, \Intd \lambda &= 0 \,  \qquad\qquad (r>R) \, , \label{EqOuterPsi}
	\end{align}
\end{subequations}
where the expression of the known radial function~$f_k(r)$ is given by Eq.~\eqref{fk}.

A formal solution of these types of dual integral equations with Bessel function kernels was first derived by Titchmarsh~\cite{titchmarsh48book}. The solution procedure involves the theory of Mellin transforms~\cite{tranter51}, but this method presents some mathematical difficulty.
In some particular cases, Sneddon and later Copson have shown that the dual integral equations problem can be reduced to classical Abel integral equations upon suitable substitutions~\cite{sneddon60, copson61}.
An analogous resolution approach has recently been employed by some of us to compute the creeping flow field due to a Stokeslet near a finite-sized elastic interface~\cite{daddi19jpsj}.

Following the recipes by Copson~\cite{copson61}, we write the solution of the dual integral equations in terms of unknown functions $\hat{\psi}_k$ as
\begin{equation}
	\psi_k(\lambda) = \lambda^{\halb} \int_0^R \hat{\psi}_k (t) J_{k-\halb} (\lambda t) \, \Intd t \, .
	\label{psiKInit}
\end{equation}
The latter equation can be rewritten as
\begin{equation}
	\psi_k(\lambda) = \lambda^{-\halb} \int_0^R \hat{\psi}_k (t) t^{-\left(k+\halb\right)}
	\frac{\Intd}{\Intd t} \left( t^{k+\halb} J_{k+\halb} (\lambda t) \right) \, \Intd t \, .
	\label{psiKORIG}
\end{equation}
Defining 
\begin{equation}
	\hat{\Psi}_k (t) = t^{k+\halb} \frac{\Intd}{\Intd t} \left( \hat{\psi}_k(t) \, t^{-\left(k+\halb\right)} \right)  
\end{equation}
and assuming that $\lim\limits_{t\to 0^+} t^{k+\halb} \, \hat{\psi}_k(t) = 0$, Eq.~\eqref{psiKORIG} takes the final form after integrating by parts 
\begin{equation}
	\psi_k(\lambda) = \lambda^{-\halb} \hat{\psi}_k(R)J_{k+\halb} (\lambda R) - \int_0^R \lambda^{-\halb} \hat{\Psi}_k (t) J_{k+\halb} (\lambda t) \, \Intd t  \, . \label{psikAfterIntByParts}
\end{equation}

By making use of the following identity given by Watson~\cite{watson95},
\begin{equation}
	\int_0^\infty \lambda^{1+q-p} J_p(a\lambda) J_q(b\lambda) \Intd \lambda =
	\frac{b^q H(a-b)}{a^p \Gamma \left(p-q\right)} \left(\frac{a^2-b^2}{2}\right)^{p-q-1} ,
	\label{watson} 
\end{equation}
with~$H(\cdot)$ denoting the Heaviside step function and $\Gamma(\cdot)$ Euler's Gamma function~\cite{abramowitz72}, it can readily be checked that Eq.~\eqref{psikAfterIntByParts} satisfies  the outer integral equation~\eqref{EqOuterPsi} upon inversion of the order of integration.

Substitution of Eq.~\eqref{psiKInit} into the inner problem stated by Eq.~\eqref{EqInnerPsi} leads to
\begin{equation}
	\left(\frac{2}{\pi}\right)^{\halb} r^{-k} 
	\int_0^r t^{k-\halb} \hat{\psi}_k (t) \left( r^2-t^2 \right)^{-\halb} \, \Intd t = f_k(r) \, . \label{psiAsAbelEqn}
\end{equation}
The latter result is obtained after making use of Eq.~\eqref{watson} and noting that~$\Gamma(1/2)=\pi^{1/2}$.

Equation~\eqref{psiAsAbelEqn} is a Volterra integral equation of the first kind~\cite{carleman21, smithies58, anderssen80}, which can be transformed into an Abel integral equation.
The latter admits a unique solution if~$f_k(r)$ is a continuously-differentiable function~\cite{whittaker96, carleman22, tamarkin30}.
More generally, the solution of the integral equation 
\begin{equation}
	\int_0^r g(t) \left(r^2-t^2\right)^{-\alpha} \, \Intd t = h(r) \,  \quad (0<r<R) \, , 
\end{equation}
defined for~$0<\alpha<1$ is of the form
\begin{equation}
	g(r) = \frac{2}{\pi} \, \sin(\alpha\pi) \, \frac{\Intd}{\Intd r} 
	\int_0^r h(t) (r^2-t^2)^{\alpha-1} t \, \Intd t \, .
\end{equation}
Hence, the solution of Eq.~\eqref{psiAsAbelEqn} is obtained as
\begin{equation}
	\hat{\psi}_k (r) = \left(\frac{2}{\pi}\right)^{\halb} r^{-k+\halb} \frac{\Intd }{\Intd r}
	\int_0^r t^{k+1} f_k(t) \left(r^2-t^2\right)^{-\halb} \, \Intd t \, .
\end{equation}

Finally, we get
\begin{equation}
		\hat{\psi}_k (t) = -2\left(\frac{2}{\pi}\right)^{\halb} 
		\frac{h^2 t^{3/2}}{\left(t^2+h^2\right)^2} \, \delta_{k1} \, ,
\end{equation}
and the sought-for functions $\psi_k$ are obtained from Eq.~\eqref{psiKInit}. We note that
\begin{equation}
	J_{\halb} (\lambda t) = \left( \frac{2}{\pi \lambda t} \right)^{\halb} \sin(\lambda t) \, .
\end{equation}

\subsection{Solution for~$\pi_k(\lambda)$ and~$\omega_k(\lambda)$}

We next consider the system of dual integral equations for the unknown wavenumber-dependent functions~$\pi_k(\lambda)$ and~$\omega_k(\lambda)$, which is given for the inner region $r<R$  by
\begin{subequations}\label{DualIntEqsPiUndOmegaInner}
	\begin{align}
		\int_0^\infty 
		\lambda \left( \pi_k(\lambda) - \omega_k(\lambda) \right) J_{k+1}(\lambda r) \, \Intd \lambda &=  f^-_k (r) \, , \label{EqInner1} \\
		\int_0^\infty 
		\lambda \left( \pi_k(\lambda) + \omega_k(\lambda) \right) J_{k-1}(\lambda r) \, \Intd \lambda &= f^+_k(r) \, ,  \label{EqInner2}
	\end{align}
\end{subequations}
 while for the outer region $r>R$ we have
\begin{subequations}\label{DualIntEqsPiUndOmegaOuter}
	\begin{align}
		\int_0^\infty 
		\lambda^2 \left( 2\pi_k(\lambda) - \omega_k(\lambda) \right) J_{k+1}(\lambda r) \, \Intd \lambda &= 0 \, , \label{EqOuter1} \\
		\int_0^\infty 
		\lambda^2 \left( 2\pi_k(\lambda) + \omega_k(\lambda) \right) J_{k-1}(\lambda r) \, \Intd \lambda &= 0 \, , \label{EqOuter2} 
	\end{align}
\end{subequations}
 where~$f_k^-(r)$ and~$f_k^+(r)$ are defined by Eqs.~\eqref{fkMinus} and~\eqref{fkPlus}, respectively.
Notably, the integral equations for~$\pi_k(\lambda)$ and~$\omega_k(\lambda)$ cannot be decoupled, rendering the problem more challenging. 
Consequently, we have to solve simultaneously for these two quantities.

In order to satisfy the equations for the outer region, we follow the approach outlined by Copson~\cite{copson61} and seek solutions of the integral form 
\begin{subequations}\label{formSolutionPiUNDOmega}
	\begin{align}
		2\pi_k(\lambda)-\omega_k(\lambda) &= \lambda^{-\halb} \int_0^R \hat{\xi}_k^{-}(t)J_{k+\halb}(\lambda t) \, \Intd t \, , \\
		2\pi_k(\lambda)+\omega_k(\lambda) &= \lambda^{-\halb} \int_0^R \hat{\xi}_k^{+}(t)J_{k-\frac{3}{2}}(\lambda t) \, \Intd t \, ,
	\end{align}
\end{subequations}
where~$\hat{\xi}_k^{-}(t)$ and~$\hat{\xi}_k^{+}(t)$ are unknown functions to be determined by satisfying the mixed dual integral equations in the inner region.

\begin{widetext}

Performing integration by parts in Eqs.~\eqref{formSolutionPiUNDOmega} yields 
\begin{subequations}\label{IntByPartsPiUNDOmega}
	\begin{align}
		2\pi_k(\lambda)&-\omega_k(\lambda) = \lambda^{-\frac{3}{2}}
		\Bigg( \hat{\xi}_k^{-}(R)J_{k+\frac{3}{2}} (\lambda R) -\int_0^R t^{k+\frac{3}{2}} \frac{\Intd}{\Intd t}
		\left( \hat{\xi}_k^{-}(t) \, t^{-\left(k+\frac{3}{2}\right)} \right)
		J_{k+\frac{3}{2}} (\lambda t) \, \Intd t
		\Bigg) , \\
		2\pi_k(\lambda)&+\omega_k(\lambda) = \lambda^{-\frac{3}{2}}
		\Bigg( \hat{\xi}_k^{+}(R)J_{k-\frac{1}{2}} (\lambda R) 
		-\int_0^R t^{k-\frac{1}{2}} \frac{\Intd}{\Intd t} 
		\left( \hat{\xi}_k^{+}(t) \, t^{-\left(k-\frac{1}{2}\right)} \right) J_{k-\frac{1}{2}} (\lambda t) \, \Intd t
		\Bigg)  , 
	\end{align}
\end{subequations}
where we have assumed that $\lim\limits_{t\to 0^+} t^{k+\frac{3}{2}} \hat{\xi}_k^{-}(t) = \lim\limits_{t\to 0^+} t^{k-\frac{1}{2}} \hat{\xi}_k^{+}(t) = 0$.

By making use of the identity by Watson given by Eq.~\eqref{watson} and inverting the order of integration, it can readily be verified that Eqs.~\eqref{IntByPartsPiUNDOmega} satisfy the outer problem stated by Eqs.~\eqref{DualIntEqsPiUndOmegaOuter}.

Solving Eqs.~\eqref{formSolutionPiUNDOmega} for~$\pi_k(\lambda)$ and~$\omega_k(\lambda)$ yields
\begin{subequations}\label{PiUndOmegaK}
	\begin{align}
		\pi_k(\lambda) &= \frac{\lambda^{-\halb}}{4} \int_0^R
		\left( \hat{\xi}_k^{-}(t)J_{k+\halb}(\lambda t) + \hat{\xi}_k^{+}(t)J_{k-\frac{3}{2}} (\lambda t) \right) \Intd t \, , \\
		\omega_k(\lambda) &= \frac{\lambda^{-\halb}}{2} \int_0^R
		\left( -\hat{\xi}_k^{-}(t)J_{k+\halb}(\lambda t) + \hat{\xi}_k^{+}(t)J_{k-\frac{3}{2}} (\lambda t) \right) \Intd t \, .
	\end{align}
\end{subequations}

Next, substituting Eqs.~\eqref{PiUndOmegaK} into Eqs.~\eqref{DualIntEqsPiUndOmegaInner} and interchanging the order of integration yields
\begin{subequations} \label{innerProblemK}
	\begin{align}
		\int_0^R \left( \frac{3}{4} \, \mathcal{K}_{k+\halb}^{k+1} (r,t) \hat{\xi}_k^{-}(t)  
		- \frac{1}{4} \, \mathcal{K}_{k-\frac{3}{2}}^{k+1} (r,t) \hat{\xi}_k^{+}(t) \right) \Intd t 
		= f_k^- (r) \, ,  \\
		 \int_0^R \left( -\frac{1}{4} \, \mathcal{K}_{k+\halb}^{k-1} (r,t) \hat{\xi}_k^{-}(t) 
		+ \frac{3}{4} \, \mathcal{K}_{k-\frac{3}{2}}^{k-1} (r,t) \hat{\xi}_k^{+}(t) \right) \Intd t 
		= f_k^+ (r) \, , 
	\end{align}
\end{subequations}
where we have defined the kernel functions
\begin{equation}
	\mathcal{K}_{p}^{q} (r,t) = \int_0^\infty \lambda^\halb J_p(\lambda t) J_q (\lambda r) \, 
	\Intd \lambda \, ,
\end{equation}
where~$p=-\frac{3}{2}, -\halb, \halb, \frac{3}{2}, \dots$ and~$q=-1, 0, 1, 2, \dots$.
It can be shown that~$\mathcal{K}_{p}^{q} (r,t)$ is always convergent in the range of definition of~$p$ and~$q$.

In particular, the kernel functions appearing in Eqs.~\eqref{innerProblemK} are explicitly given by
\begin{subequations} \label{KFunktionen}
	\begin{align}
		\mathcal{K}_{k+\halb}^{k+1} (r,t)
		&= 
		\left(\tfrac{2}{\pi}\right)^\halb r^{-(k+1)} t^{k+\halb} \left(r^2-t^2\right)^{-\halb}
		H(r-t) \, , \\
		\mathcal{K}_{k-\frac{3}{2}}^{k+1} (r,t)
		&=
		\left(\tfrac{2}{\pi}\right)^\halb r^{-(k+1)} t^{k-\frac{3}{2}}
		\left( 
		(2k-1) \left( r^2-t^2 \right)^\halb - t^2 \left(r^2-t^2\right)^{-\halb}
		\right) H(r-t) \, , \\
		\mathcal{K}_{k+\halb}^{k-1} (r,t)
		&=	
		\left(\tfrac{2}{\pi}\right)^\halb r^{-(k+2)} t^{k+\halb} 
		\Lambda_k \left( \tfrac{t}{r} \right) H(r-t)
		+ 2^\halb r^{k-1} t^{-\left(k+\halb\right)} \Gamma \left( k+\tfrac{1}{2} \right) \Gamma (k)^{-1} H(t-r) \, , \\
		\mathcal{K}_{k-\frac{3}{2}}^{k-1} (r,t)
		&=
		\left(\tfrac{2}{\pi}\right)^\halb r^{-(k-1)} t^{k-\frac{3}{2}} \left(r^2-t^2\right)^{-\halb} H(r-t) \, .
	\end{align}
\end{subequations}
In addition to that, we have defined the series function
\begin{equation}
	\Lambda_k(x) = \frac{\left( 1 + 2k \left( 1-x^2 \right) \right)
		\delta_k^{+} (x) - 2(k+1) \, \delta_k^{-} (x)}{(1+2k) (1-x^2)} \,  \qquad
	|x| < 1 \, , 
\end{equation}
where 
\begin{equation}
	\delta_k^\pm (x) = {}_2 F_1 \left( \pm \frac{1}{2} ,k+\frac{1}{2}; k+\frac{3}{2}; x^2 \right) ,
\end{equation}
with~${}_2 F_1$ denoting the Gauss hypergeometric functions~\cite{abramowitz72}.
We note that~$\lim\limits_{k \to 0} \Gamma(k)^{-1} = 0$.
It is worth mentioning that, for~$k \ge 1$, the series function $\Lambda_k (x)$ can be represented in terms of usual analytic functions as
\begin{equation}
	\Lambda_k(x) = \frac{2^{-(2k-1)} x^{-(2k+1)} }{(k-1)!}
	\left(\frac{(2k)!}{k!} \, \arcsin(x) + \frac{2^k}{2k+1} (1-x^2)^{-\halb} Q_k(x) \right) \, , 
\end{equation}
where
\begin{equation}
	Q_k(x) = (2k+1)!! \left( -x + \sum_{i=1}^{k-1} 2^{2i-1} \frac{ i!(i-1)!}{(2i+1)!} \, x^{2i+1}\right) 
\end{equation}
is a polynomial in powers of~$x$ of degree $2k-1$. Here, $k!!$ denotes the double factorial of~$k$ (also sometimes called semifactorial), defined as the product of all the preceding integers that have the same parity~\cite{meserve48}.
For~$k=0$, we note that~$\Lambda_0(x) = -\left(1-x^2\right)^{-\halb}$.

By inserting Eqs.~\eqref{KFunktionen} into Eqs.~\eqref{innerProblemK} and using the substitution
\begin{align}
	\hat{\chi}_k^\pm (t) = \frac{1}{4} \left(\frac{2}{\pi t}\right)^\halb \hat{\xi}_k^\pm (t) \, , 
	\label{chiVersusXi}
\end{align}
the integral equations for the inner problem can be presented in the final form
\begin{subequations} \label{InnerProblemSihammiert}
	\begin{align}
		r^{-(k+1)}\int_0^r
		\left( 3 \left(r^2-t^2\right)^{-\halb} \hat{\chi}_k^{-}(t)
		- \left( t^{-2} (2k-1) \left( r^2-t^2 \right)^\halb - \left(r^2-t^2\right)^{-\halb} \right) \hat{\chi}_k^{+}(t)
		\right) t^{k+1} \, \Intd t &= f_k^-(r) \, ,  \\
		r^{-k} \int_0^r
		\left( - \Lambda_k  \left(\tfrac{t}{r}\right) r^{-2} \hat{\chi}_k^{-}(t) + 3r t^{-2} \left(r^2-t^2\right)^{-\halb} \hat{\chi}_k^{+}(t) \right) t^{k+1} \, \Intd t 
		-\pi^\halb \Gamma \left( k+\tfrac{1}{2} \right) \Gamma (k)^{-1} \, r^{k-1}
		\int_r^R t^{-k} \hat{\chi}_k^{-}(t) \, \Intd t &= f_k^+(r) \, .
	\end{align}	
\end{subequations}

For the solution of Eqs.~\eqref{InnerProblemSihammiert} we use the standard series expansion technique~\cite{goldfine77, kanwal89, ren99}. 
The method consists of writing the solutions of the integral equations as power series expansions with unknown coefficients of the form
\begin{equation}
	\hat{\chi}_k^\pm(t) = \sum_{p\in \mathbb{Z}} a_{k,p}^\pm t^p 
\end{equation}
and solving for the unknown coefficients by identification of terms of the same power when comparing with the series expansion of the known function.
Since $f_k^\pm (k) = 0$ for~$k \ne 1$, we set $a_{k,p}^\pm = 0, \forall p \in \mathbb{Z}$ in this case.
As a result, it follows that~$\hat{\chi}_k^\pm (t) = 0$ for~$k \ne 1$.
As for~$k=1$, knowing that~$\Lambda_1(x) = x^{-3} \arcsin (x) - x^{-2}\left(1-x^2\right)^{-\halb}$, Eqs.~\eqref{InnerProblemSihammiert} yield
\begin{subequations}\label{InnerProblemSihammiertForK1}
	\begin{align}
		\int_0^r \left( 3t^2 \left(r^2-t^2\right)^{-\halb} \hat{\chi}_1^-(t) 
		+ \left( t^2 \left(r^2-t^2\right)^{-\halb} - \left(r^2-t^2\right)^{\halb}  \right) \hat{\chi}_1^+(t) \right) \Intd t &= \frac{r^4}{\left( r^2+h^2 \right)^\frac{3}{2}} \, , \\
		\int_0^r \left( \left(\left(r^2-t^2\right)^{-\halb}-t^{-1} \arcsin \left(\frac{t}{r}\right)  \right) \hat{\chi}_1^-(t)+3 \left(r^2-t^2\right)^{-\halb} \hat{\chi}_1^+(t) \right) \Intd t 
		-\frac{\pi}{2} \int_r^R t^{-1} \hat{\chi}_1^-(t) \, \Intd t
		&= -\frac{3r^2+2h^2}{\left(r^2+h^2\right)^\frac{3}{2}} \, .
	\end{align}
\end{subequations}

We express the solution as an even function of~$t$ as
\begin{equation}
	\hat{\chi}_1^\pm(t) = \frac{1}{\pi h} \sum_{p \ge 0} c_p^\pm \left(\frac{t}{h}\right)^{2p} \, ,
	\label{chi1FormSolution} 
\end{equation}
where~$t<h$ and $c_p^\pm$ are constants to be determined by identification.
Substituting Eq.~\eqref{chi1FormSolution} into Eqs.~\eqref{InnerProblemSihammiertForK1} and expanding the right-hand sides in Taylor series about the origin yields
\begin{subequations}
	\begin{align}
		\frac{3}{4} \, c_0^- \rho^2 + \frac{9c_1^-+2c_1^+}{16} \, \rho^4 + \frac{15c_2^-+4c_2^+}{32} \, \rho^6
		+\frac{15\left(7c_3^-+2c_3^+\right)}{256} \, \rho^8 + \cdots
		&=
		\rho^4 - \frac{3}{2} \, \rho^6 + \frac{15}{8} \, \rho^8 - \cdots  , \notag \\
		\frac{3}{2} \, c_0^+ - \frac{1}{4} \sum_{p \ge 1} \frac{c_p^-}{p} \left(\frac{R}{h}\right)^{2p}
		+\frac{3\left(c_1^- +2c_1^+\right)}{8} \, \rho^2 + \frac{3\left(5c_2^-+12c_2^+\right)}{64} \, \rho^4
		+\frac{5\left(7c_3^-+18c_3^+\right)}{192} \, \rho^6 + \cdots 
		&= -2 + \frac{3}{4} \, \rho^4 - \frac{5}{4} \, \rho^6 + \cdots \notag ,
	\end{align}
\end{subequations}
\end{widetext}
where we have defined~$\rho=r/h$.
By identification, we obtain the series coefficients
\begin{subequations}\label{ExprSeriesCoeffs}
	\begin{align}
		c_p^- &= (-1)^{p+1} 2p \,  \qquad\qquad (p \ge 0) \, , \\
		c_p^+ &= (-1)^{p} (2p-1) \,  \,\,\,\qquad (p \ge 1) \, , \\
		c_0^+ &= -\frac{3R^2 + 4h^2}{3\left( R^2 + h^2 \right)} \, .
	\end{align}
\end{subequations}

By inserting Eqs.~\eqref{ExprSeriesCoeffs} into Eq.~\eqref{chi1FormSolution} and evaluating the infinite series, we obtain 
\begin{subequations}
	\begin{align}
		\hat{\chi}_1^-(t) &= \frac{2h t^2}{\pi \left(t^2 + h^2\right)^2} \, , \\
		\hat{\chi}_1^+(t) &= \frac{1}{\pi h} \left( \frac{t^2 \left(t^2 - h^2\right)}{\left(t^2 + h^2\right)^2} - \frac{3R^2 + 4h^2}{3 \left( R^2 + h^2 \right)} \right) \, .
	\end{align}
\end{subequations}

Finally, the expression of~$\hat{\xi}_1^\pm(t)$ can be obtained from Eq.~\eqref{chiVersusXi} and the solution for~$\pi_1(\lambda)$ and~$\omega_1(\lambda)$ can then be calculated from Eqs.~\eqref{PiUndOmegaK}.
We note that
\begin{subequations}
	\begin{align}
		J_\frac{3}{2} (\lambda t) &= \left( \frac{2}{\pi \lambda t} \right)^\halb 
		\left( \frac{\sin(\lambda t)}{\lambda t} - \cos(\lambda t) \right) \, , \\
		J_{-\halb} (\lambda t) &= \left( \frac{2}{\pi \lambda t} \right)^\halb \cos (\lambda t) \, .
	\end{align}
\end{subequations}

We have checked the correctness of our derived solution by direct comparison of the left-hand sides of Eqs.~\eqref{InnerProblemFinalized} using our expressions for~$\pi_k(\lambda)$, $\omega_k(\lambda)$, and $\psi_k(\lambda)$.

\subsection{Solution for~$R\to\infty$}

Having derived the solution of the flow problem for a Stokeslet force near a finite-sized disk, we next check the correctness of the solution for an infinitely extended disk that is fixed in space.
By taking the upper limit of integration to infinity in Eqs.~\eqref{psiKInit} and~\eqref{PiUndOmegaK}, we obtain
\begin{subequations}\label{psipiomega}
	\begin{align}
		\pi_k(\lambda) &= \frac{\lambda h - 1}{\lambda} \, e^{-\lambda h} \, \delta_{k1} \, , \\
		\omega_k (\lambda) &= -\frac{2}{\lambda} \, e^{-\lambda h} \, \delta_{k1} \, , \\
		\psi_k(\lambda) &= -\lambda h e^{-\lambda h} \, \delta_{k1} \, .
	\end{align}
\end{subequations}

It is worth mentioning that these solutions can likewise be obtained by applying inverse Hankel transforms~\cite{davies12book} to Eqs.~\eqref{InnerProblemFinalized}.
Then,
\begin{subequations}
	\begin{align}
		\pi_k(\lambda) - \omega_k(\lambda) &=
		\int_0^\infty r f_k^-(r) J_{k+1} (\lambda r) \, \Intd r \, , \\
		\pi_k(\lambda) + \omega_k(\lambda) &=
		\int_0^\infty r f_k^+(r) J_{k-1} (\lambda r) \, \Intd r \, , \\
		\psi_k(\lambda) &= \lambda \int_0^\infty r f_k(r) J_k(\lambda r) \, \Intd r \, .
	\end{align}
\end{subequations}

By inserting Eqs.~\eqref{psipiomega} into Eq.~\eqref{PiPsiOmega}, the harmonic functions~$\Pi_k^\pm$, $\Psi_k^\pm$, and~$\Omega_k^\pm$ are expressed in the limit~$R\to\infty$ by
\begin{subequations}
	\begin{align}
		\Pi_k^\pm  &= \frac{-F}{8\pi\eta} \frac{\cos\phi}{r} \left( 2h \pm z -s_\pm - \frac{h(h \pm z)}{s_\pm} \right) \delta_{k1} \, , \\
		\Psi_k^\pm &= \frac{F}{8\pi\eta}\frac{hr \cos\phi}{s_\pm^3} \,\delta_{k1} \, , \\
		\Omega_k^\pm &= \frac{-F}{4\pi\eta} \frac{\sin\phi}{r} \left( h \pm z - s_\pm \right) \delta_{k1} \, , 
	\end{align}
\end{subequations}
where we have defined
\begin{equation}
	s_\pm = \left( r^2 + (h \pm z)^2 \right)^\halb \, .
\end{equation}

This solution is in full agreement with the familiar solution originally obtained by Lorentz~\cite{lorentz07} for a Stokeslet exerted tangent to an infinitely extended stationary plane boundary using the method of reflections. This solution has later been derived and analyzed by Blake using a Fourier transform technique~\cite{blake71,skultety20}.

\input{main.bbl}

\end{document}

%% file: commands.tex
\newcommand{\vect}[1]{\boldsymbol{#1}}

\newcommand{\bNabla}{\boldsymbol{\nabla}}

\newcommand{\R}{\vect{r}}

\newcommand{\Intd}{\mathrm{d}}

\newcommand{\halb}{\frac{1}{2}}

%% file: main.bbl
%